\documentclass[epp]{agutex}

\usepackage{multirow,color}
\usepackage{bbding}
\usepackage{amssymb}
\usepackage{ulem}
\usepackage{verbatim}
\usepackage{url}

\usepackage{graphicx}
\usepackage[displaymath]{lineno}

\def\gsim{\;\lower4pt\hbox{${\buildrel\displaystyle >\over\sim}$}\;}
\def\lsim{\;\lower4pt\hbox{${\buildrel\displaystyle <\over\sim}$}\;}
\def\grls{\;\lower4pt\hbox{${\buildrel\displaystyle >\over <}$}\;}

\newcommand{\mod}[1]{{\color[rgb]{0,0,0} #1}}
\newcommand{\del}[1]{{\color[rgb]{0,0,1} \sout{#1}}}
\renewcommand{\del}[1]{{}}

\volume{0}
\authorrunninghead{WANG ET AL.}
\titlerunninghead{Mars Orbiter Magnetometer of Tianwen-1}

\begin{document}

%\title{First Results of the Mars Orbiter Magnetometer onboard Tianwen-1}
\title{The Mars Orbiter Magnetometer of Tianwen-1: In-flight Performance and First Science Results}

\author{Yuming Wang,$^{1,2,*}$ Tielong Zhang,$^{2,3}$ 
Guoqiang Wang,$^{4}$ Sudong Xiao,$^{4}$ Zhuxuan Zou,$^{1,2}$ Long Cheng,$^{1,2}$ 
Zonghao Pan,$^{1,2}$ Kai Liu,$^{1,2}$ Xinjun Hao,$^{1,2}$ Yiren Li,$^{1,2}$ Manming Chen,$^{1,2}$ 
Zhoubin Zhang,$^{5}$ Wei Yan,$^{5}$
Zhenpeng Su,$^{1,2}$ Zhiyong Wu,$^{1,2}$ Chenglong Shen,$^{1,2}$ 
Yutian Chi,$^{6}$ Mengjiao Xu,$^{6}$ Jingnan Guo,$^{1,2}$ and Yang Du$^{7}$}

\affil{$^1$ School of Earth and Space Sciences/Deep Space Exploration Laboratory, University of Science and Technology of China, Hefei 230026, China}

\affil{$^2$ CAS Center for Excellence in Comparative Planetology/CAS Key Laboratory of Geospace Environment/Mengcheng National Geophysical Observatory, University of Science and Technology of China, Hefei 230026, China}

\affil{$^3$ Space Research Institute, Austrian Academy of Sciences, Graz, Austria}

\affil{$^4$ Institute of Space Science and Applied Technology, Harbin Institute of Technology, Shenzhen, China}

\affil{$^5$ National Astronomical Observatories, Chinese Academy of Sciences, Beijing, China}

\affil{$^6$ Institute of Deep Space Sciences, Deep Space Exploration Laboratory, Hefei 230026, China}

\affil{$^7$ Shanghai Institute of Satellite Engineering, Shanghai, China}

\affil{$^*$ Corresponding author, Email: ymwang@ustc.edu.cn}

\begin{abstract}
Mars Orbiter MAGnetometer (MOMAG) is a scientific instrument onboard the orbiter of China's first mission 
for Mars --- Tianwen-1. It started to routinely measure the magnetic field from the solar wind to magnetic 
pile-up region surrounding Mars since November 13, 2021. Here we present its in-flight performance and first science results 
based on the first one and a half months' data. By comparing with the magnetic field data in the solar wind from 
the Mars Atmosphere and Volatile EvolutioN (MAVEN), 
the magnetic field by MOMAG is at the same level in magnitude, and the same magnetic structures with the similar 
variations in three components could be found in MOMAG data. In the first one and a half months, we recognize 158 clear 
bow shock (BS) crossings from MOMAG data, whose locations statistically match well with the modeled average BS. 
We also identify 5 pairs of simultaneous BS crossings of the Tianwen-1's orbiter and MAVEN. These BS 
crossings confirm the global shape of modeled BS as well as the south-north asymmetry of the Martian BS. \mod{Two presented 
cases in this paper suggest} that the BS is probably more dynamic at flank than near the nose.
So far, MOMAG performs well, and provides accurate magnetic field vectors. MOMAG is 
continuously scanning the magnetic field surrounding Mars. These measurements complemented by observations from 
MAVEN will undoubtedly advance our understanding of the plasma environment of Mars.
\end{abstract}

\begin{article}

\section{Introduction}

Tianwen-1 is the first mission of China to explore and study Mars from its space environment to the 
surface~\citep{WanW_etal_2020, ZouY_etal_2021}. 
It consists of an orbiter,
a lander and a rover, called Zhurong. Mars Orbiter MAGnetometer (MOMAG) is one of the scientific instruments onboard 
the orbiter\citep{LiuK_etal_2020}. It investigates the magnetic field environment of Mars by measuring the local vector 
magnetic field, and therefore provides some key information for the understanding of the history and evolution of Mars.

The magnetic field surrounding Mars has two sources. One is the dynamic magnetic field resulted from the coupling between 
the solar wind and the Martian ionosphere, and the other is the static crustal magnetic 
field of Mars itself. Since Mars has no global intrinsic magnetic field, the solar wind carrying interplanetary magnetic field 
directly interacts with Martian ionosphere, and forms the bow shock (BS) and induced magnetosphere, which consists of magnetic
pileup region (MPR) and wake region\citep[e.g.,][]{Bertucci_etal_2004, Brain_etal_2006}. Between the BS and MPR, there is magnetosheath 
separated from MPR with the magnetic pileup boundary (MPB)\citep[e.g.,][]{Mazelle_etal_2004}.
The magnetic field in these regions influenced by solar wind is highly dynamic. 
%\quest{The static crustal magnetic field ...}

Escape of ions in Martian atmosphere is one of the core science issues of Tianwen-1, and is closely related to the magnetic environment.
For example, the strong static crustal magnetic field on the southern hemisphere may reach upto a high altitude and 
reconnect with interplanetary magnetic field, causing the escape of ions\citep{Brain_etal_2015}, just like the 
behavior of Venus\citep{ZhangT_etal_2012}. 
Besides, various waves in the ionosphere may heat particles, causing ion outflow\citep{Ergun_etal_2006}, and when these heated 
ions transport outside the MPB, they will interact with magnetic field carried by solar wind stream to further generate
ion cyclotron wave, boosting the escape of the ions\citep{Russell_Blanco-Cano_2007}. The escape rate during storm times will be one
to two orders higher than that during quite time\citep{Jakosky_etal_2015a}.

\begin{figure*}[h]
\begin{center}
\includegraphics[width=\hsize]{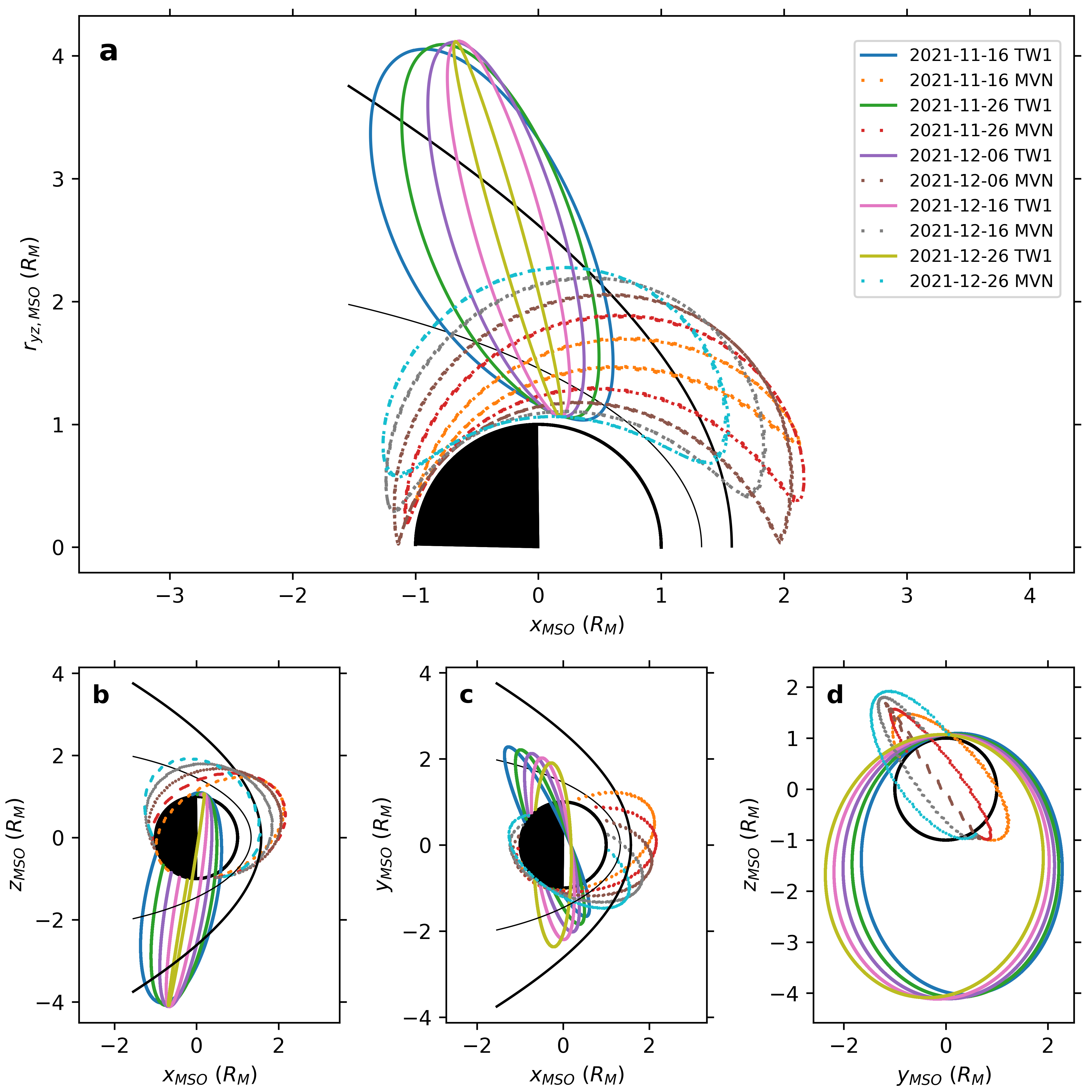}
	\caption{The orbits of Tianwen-1's orbiter (solid lines) and MAVEN (dashed lines) 
during November 13 -- December 31, 2021. The modeled Martian bow shock and MPB~\citep{Edberg_etal_2008} 
are indicated 
by thick and thin black lines, respectively.
}\label{fig:orbits}
\end{center}
\end{figure*}

Tianwen-1 orbiter was running on a highly eccentric orbit with the periapsis of about 1.08 Mars radii ($r_{m}$) and 
the apoapsis of about 4.17 $r_{m}$, and the
orbital plane is highly inclined during November and December in 2021, as shown in Figure~\ref{fig:orbits}. 
During that period, the periapsis was right above the northern pole of Mars,
the apoapsis far above the southern pole in the solar wind, and the orbital period was about 7.8 hr, with 
about 50\% -- 75\% of time in solar wind. Thus, MOMAG mainly measured the magnetic field from solar
wind to the MPR on the dawn-dusk side. Later, the inclination angle of the orbit will decrease to allow the orbiter
detect the wake region of Mars. These data will help us understand the structure and
evolution of Martian magnetic field environment and provide clues for ion escape.
Since the Mars Atmosphere and Volatile EvolutioN (MAVEN, \citealt{Jakosky_etal_2015}),
which also carries a magnetometer (MAG, \citealt{Connerney_etal_2015a}), is still working, 
the successful operation of MOMAG will make us for the first time to study 
Martian magnetic field environment from two points.

In this paper, we show and analyze the data during November 13 -- December 31, 2021.
In Section 2, we will describe the basic information and the current status of MOMAG, 
and present some measured magnetic field. Then we show the first results of MOMAG about Martian BS 
with the comparison with MAVEN/MAG data in Sections~\ref{sec_bs}.
In the last section, we summarize the paper.

\section{In-flight Calibration and Performance}\label{sec_cali}

MOMAG contains two sensors mounted on a 3.19 m long boom. The outer sensor accommodates at the 
top of the boom and the inner sensor is 0.9 m away (see \citealt{LiuK_etal_2020} for details). 
%In the spacecraft coordinates, the outer and inner
%sensors are \quest{5.65 and 4.56} m, respectively, away from the center of the spacecraft. 
Since the orbiter of 
Tianwen-1 does not have magnetic cleanliness control, the boom is actually not long enough to avoid 
the contaminations of the magnetic field from the orbiter. Thus, how to remove the magnetic interference  
based on the magnetic fields measured by the two separated sensors become pivotal.

The procedure of the mitigation of the orbiter's magnetic field generally includes two steps, which 
is similar to the procedure applied on the magnetometer of Venus Express\citep{ZhangT_etal_2008, Pope_etal_2011}.
The first step is to remove the magnetic interference due to the operations of instruments.
\mod{Such interferences behave as jumps in the magnetic field. If a real discontinuity in the solar wind 
passes the spacecraft, the amplitudes of the jump at the two sensors should be the same. However, 
since the distances of the two sensors from
the instrument are different, an artificial jump will show different amplitudes at the two sensors, and
therefore could be distinguished from real jumps.}
For these artificial jumps, we use the method of \citet{Pope_etal_2011} to remove them.
The second step is to remove the static magnetic field of the orbiter and correct the offset 
of the fluxgate magnetometer. This step is mainly based on the property of Alfv\'enic waves, of which
the magnetic field almost rotates in a plane without the change of magnitude\citep{WangG_etal_2021}. 
We process the raw data of MOMAG to the level 2 (or level C in China's convention) 
data for scientific use through these steps.
Based on the data of the first several months since November 13, 2021, when MOMAG started to
formally operate, we iterate the procedure and reach the first version of the level 2 data, that
have been released at the Planet Exploration Program Scientific Data Release System 
(\url{http://202.106.152.98:8081/marsdata/}). 
The data used in this paper and in our forthcoming
papers will also be put on the official website of 
the MOMAG team at University of Science and Technology of China (USTC, \url{http://space.ustc.edu.cn/dreams/tw1_momag/}).
A complete description of the in-flight calibration procedure as well as the demonstration of the 
reliability of the calibrated data is given in the separated paper by \citet{ZouZ_etal_2022}. 

%MOMAG has the 32-Hz data near the periapsis and apoapsis and the 1-Hz data elsewhere.
%\quest{However, two issues should be noted here. The first is that the second step of the 
%current calibration is based on Alfv\'enic property which is frequent in solar wind but rare in shock sheath 
%and Martian induced magnetosphere, and therefore the data in solar wind are well calibrated but those behind 
%the BS are not. The second is that the orbiter continuously adjusts attitude through reaction wheels 
%near the periapsis to take pictures, the interference is quite different from those without attitude adjustment and
%needs additional correction which is still in development. Thus, the MOMAG data of the current 
%version should be used in caution.}

\begin{figure*}[h]
\begin{center}
\includegraphics[width=0.79\hsize]{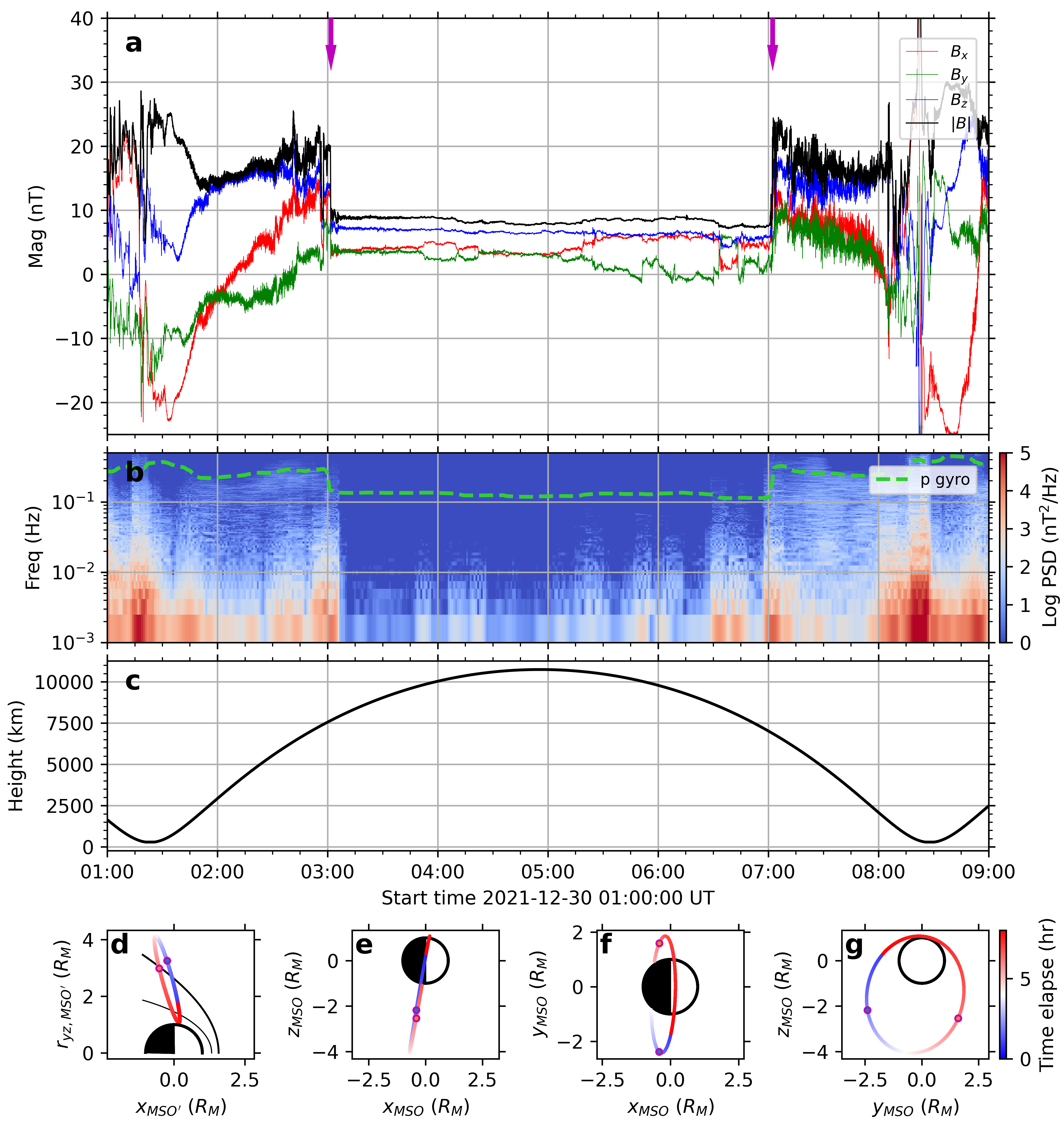}
\includegraphics[width=0.82\hsize]{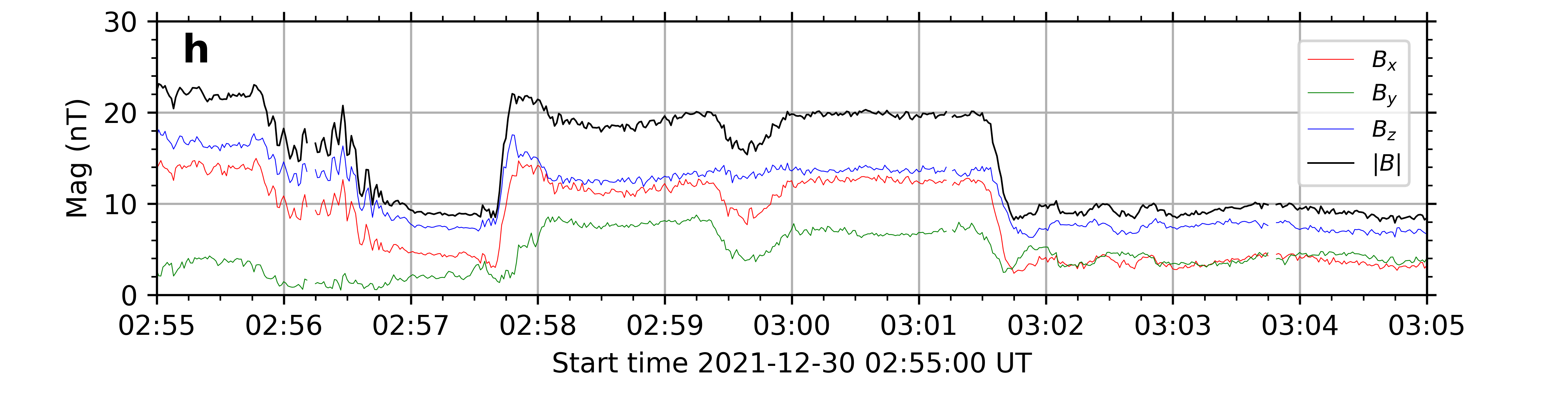}
\includegraphics[width=0.82\hsize]{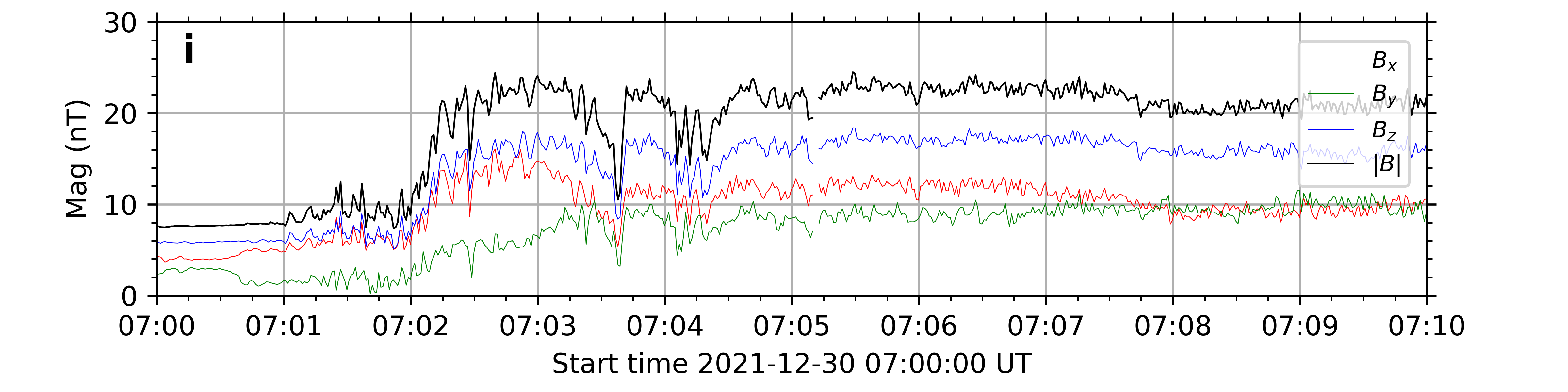}
	\caption{\footnotesize{The magnetic field measured by MOMAG during 01:00 -- 09:00 UT on December 30, 2021. 
Panel (a) shows 
the three components of the magnetic field in MSO coordinates with the total magnitude overplotted. Two purple arrows
mark the crossings of the bow shock. Panel (b) shows the power spectral density of the magnetic field fluctuations. The
green line indicates the proton's gyro frequency. Panel (c) shows the height of the Tianwen-1 orbiter, and Panel (d) -- 
(g) display its orbit in the MSO coordinates during the period of interest, in which the two circled dots mark the positions
of the bow shock crossings. \mod{Note that Panel (d) is in aberrated MSO coordinates with the modeled BS and MPB 
indicated by black lines. Panel (h) and (i) show the two BS crossings, respectively, in details.}}
}\label{fig:obstw1}
\end{center}
\end{figure*}

%The formal operation of MOMAG started on November 13, 2021 though there were sporadic tests before then. 
%The analysis of the reliability of the calibrated MOMAG data has been given in \citet{Zou_etal_2022}, we 
%just show an example here to demonstrate the data. 
Figure~\ref{fig:obstw1}a shows the magnetic field in the Mars-centered Solar Orbital (MSO) system measured by 
MOMAG during 01:00 -- 09:00 UT on 
2021 December 30. The orbiter was running in the magnetosheath before 02:55 UT, and at around \mod{03:01:40} UT the 
orbiter crossed the BS where the amplitude of the magnetic field discontinuity was more than 10 nT (Fig.\ref{fig:obstw1}h). 
After about four hours, the orbiter crossed the BS again where the amplitude of the magnetic field 
discontinuity was about 16 nT (Fig.\ref{fig:obstw1}i). 
Around 08:20 UT, the orbiter even crossed the MPB. The BS during the second 
crossing was obviously stronger than that during 
the first crossing. The reason is that the BS was compressed during the second crossing,
which can be seen from Figure~\ref{fig:obstw1}d--g that the two BS crossings were on the south,
the first crossing was outside and further away from the modeled averaged BS~\citep{Edberg_etal_2008} than the 
second BS crossing. 

\mod{If we look into the details of the first BS crossing as shown in Figure~\ref{fig:obstw1}h, 
it could be found that the orbiter crossed out the BS around 02:56:30 UT and crossed in again at 02:57:45 UT before
it finally entered the solar wind. The magnetic field changes during these preceding crossings suggest that the
BS was slightly stronger than the BS at 03:01:40 UT. This could also be explained as the compression of the BS. 
During these crossings, the orbiter was moving away from Mars as the indicated
by the color-coded orbit in Figure~\ref{fig:obstw1}d. The locations of the preceding crossings were closer to Mars than
that of the final crossing at 03:01:40 UT.}

The magnetic fields in the solar wind stayed around 9 -- 10 nT, and were much less fluctuated than those 
in the magnetosheath. Figure~\ref{fig:obstw1}b displays the power spectral density of the magnetic field. It 
is generated by using a 10-min window and 1-min running step. It can been seen that the solar wind was indeed
quiet except at very low frequency, whereas in the magnetosheath the magnetic fluctuation was enhanced. Behind
the bow shock, weak magnetic waves right below proton gyro-frequency appeared. 
In the solar wind, a small structure can be found between 05:15 and 06:35 UT. Though the total magnetic field 
only slightly enhanced, the most notable change occurred for $B_y$, which decreased from about 4 nT to zero 
twice. 
%\quest{[more analysis about the structure?]}

\begin{figure*}[h]
\begin{center}
\includegraphics[width=0.79\hsize]{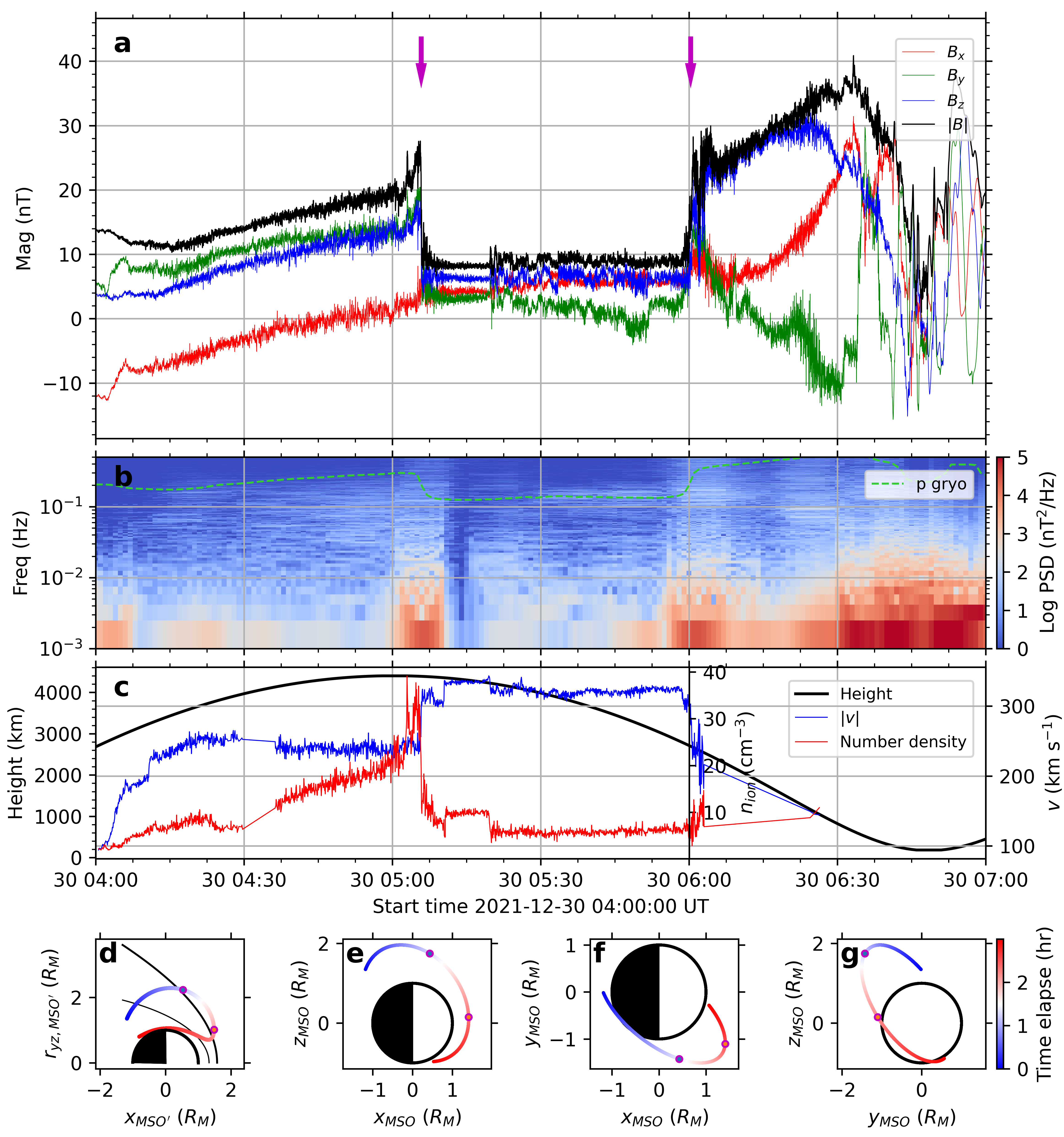}
\includegraphics[width=0.82\hsize]{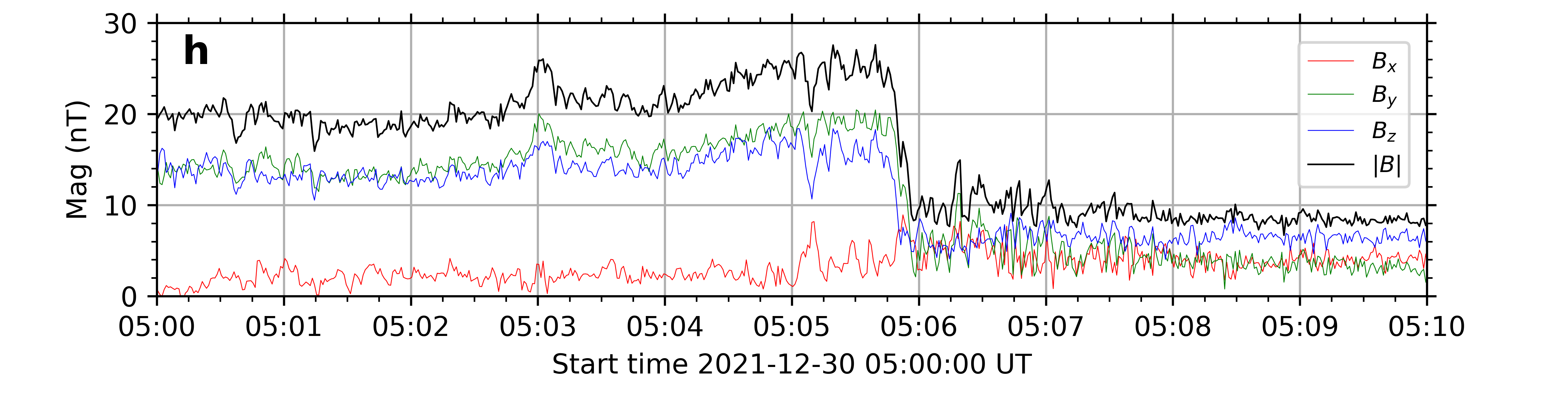}
\includegraphics[width=0.82\hsize]{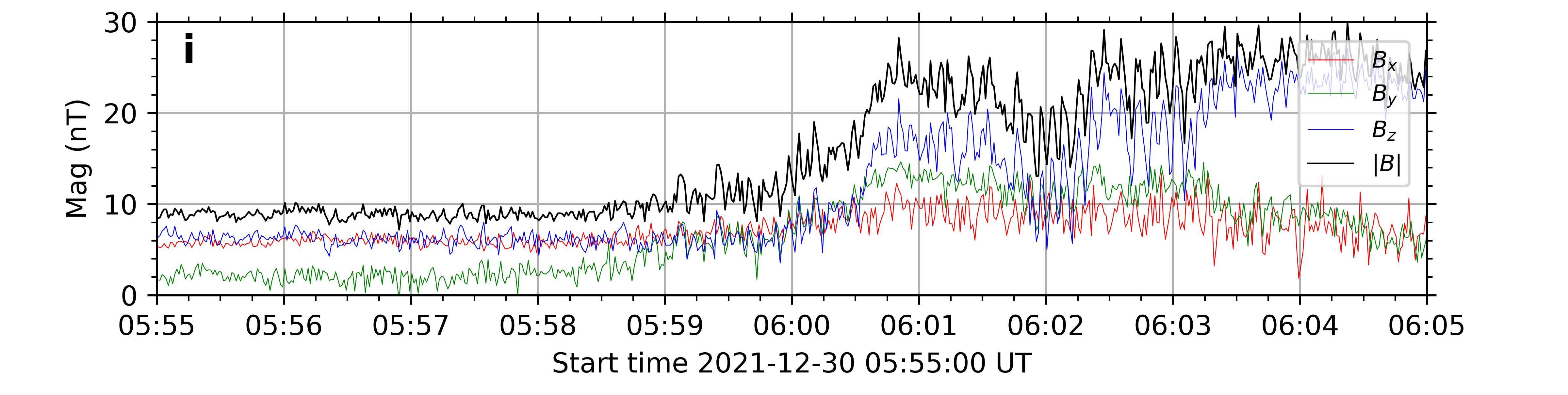}
	\caption{The magnetic field and solar wind plasma measured by MAVEN during the same period as Fig.\ref{fig:obstw1}.
In Panel (c) the solar wind speed and number density of ions are presented with the blue and red lines, respectively.
The arrangements of other panels are the same as those in Fig.\ref{fig:obstw1}.
}\label{fig:obsmvn}
\end{center}
\end{figure*}

For comparison, Figure~\ref{fig:obsmvn} shows the MAVEN measurements of the magnetic field and solar wind during 04:00
-- 07:00 UT on the same day. MAVEN had a quite different orbit, of which the orbital period was shorter than
4 hours (Fig.\ref{fig:obsmvn}c--g). Within one hour, it crossed the BS twice, but the positions of the crossings 
were both closer to the shock nose than those of Tianwen-1. Since the two crossings stay close to the same modeled BS, 
suggesting the solar wind conditions during the two crossings are almost the same, the amplitudes of their magnetic field 
discontinuities were similar. \mod{We also show the detailed BS crossings in Figure~\ref{fig:obsmvn}h and i. No multiple BS
crossings happened at MAVEN, probably suggesting the different behavior of BS at different locations.}

Though the solar wind region that MAVEN detected was far away from that by 
Tianwen-1, a similar structure could be found between 05:20 and 05:55 UT (Fig.\ref{fig:obsmvn}a), 
which corresponds to the first dip 
recorded in MOMAG. During that time, the solar wind speed was about 310 km s$^{-1}$ and the number density of protons 
was about 7 cm$^{-3}$ (see the blue and red lines in Fig.\ref{fig:obsmvn}c). 
Different from MOMAG data, the magnetic field at MAVEN was highly fluctuated with a notable 
frequency at the proton gyro frequency (Fig.\ref{fig:obsmvn}b). \mod{This might be because
MAVEN closer to Mars than Tianwen-1 when they fly in the solar wind, and the particles escaping from 
Martian atmosphere more easily interact with the solar wind and interplanetary magnetic field to generate 
such fluctuations at a closer distance. However, according to the statistical study of MAVEN 
data~\citep{Ruhunusiri_etal_2017}, such a pattern seems not to be evident and needs further validation.}

\newpage

\begin{figure*}[h]
\begin{center}
\includegraphics[width=0.485\hsize]{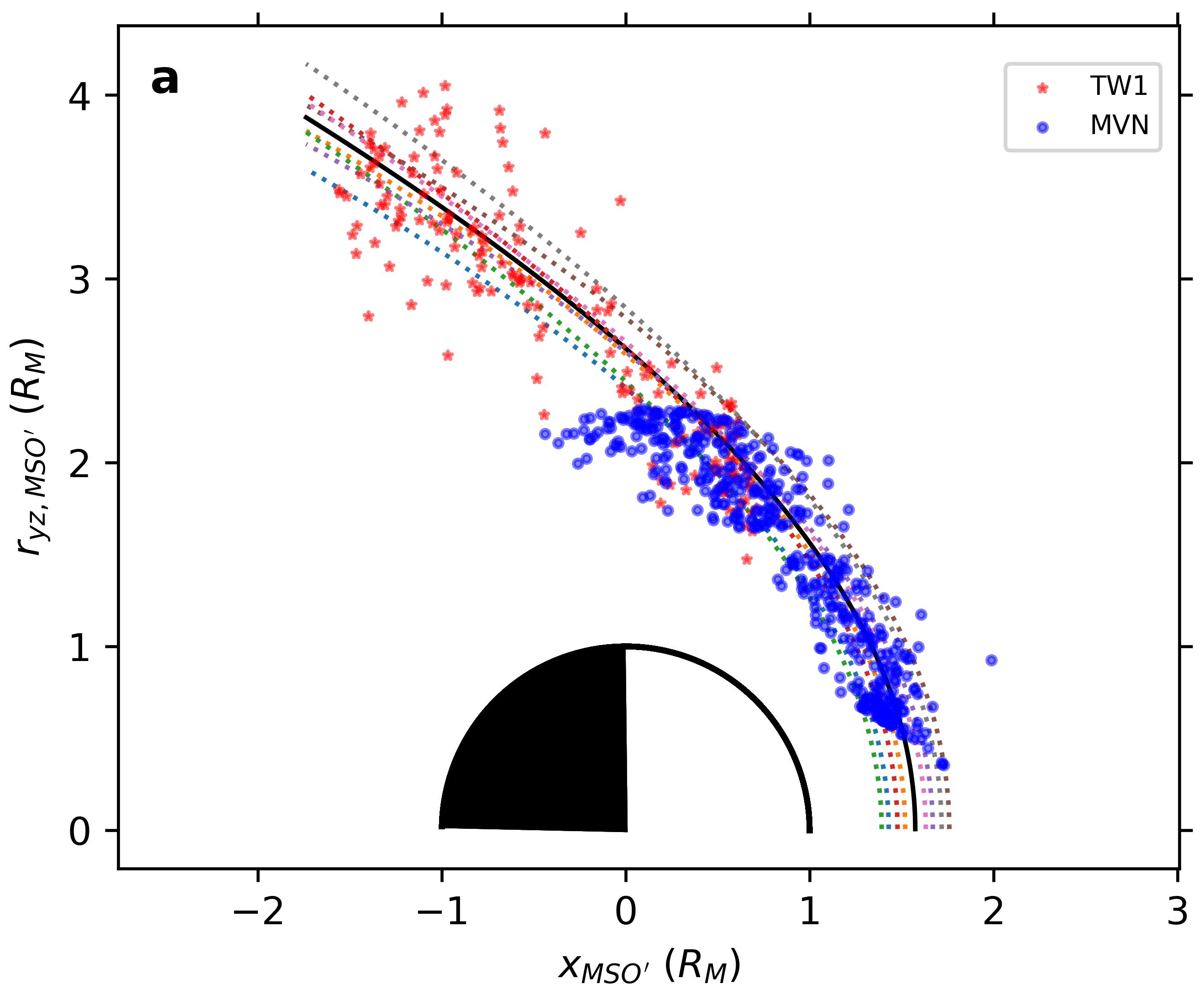}
\includegraphics[width=0.495\hsize]{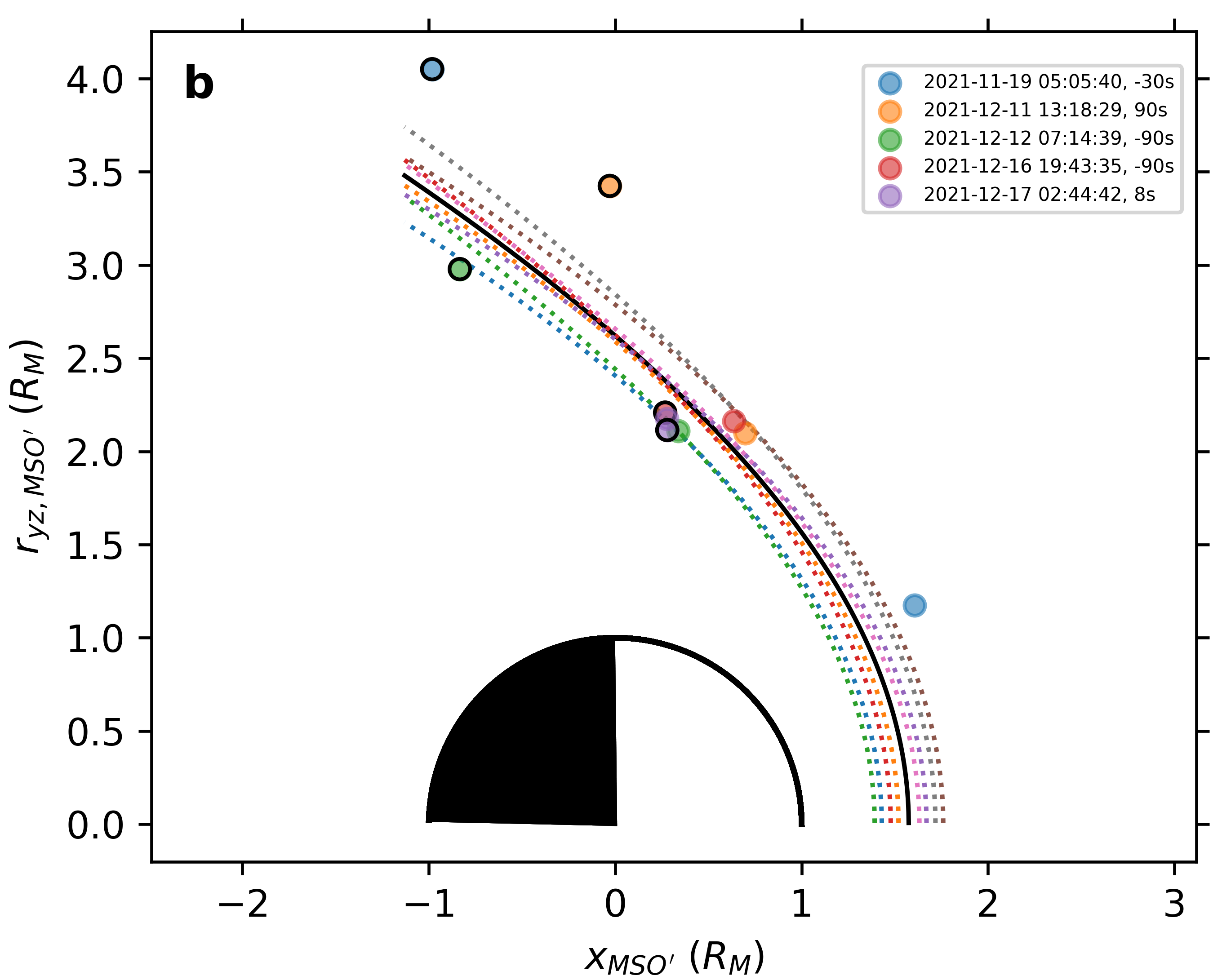}
	\caption{The bow shock (BS) crossings during the period of interest. In Panel (a), the red asterisks mean the 
BS crossings of Tianwen-1's orbiter, and the blue dots the BS crossings of MAVEN. The modeled average 
BS~\citep{Edberg_etal_2008} is displayed by the black line, and the other dashed lines display the BS when 
the uncertainties of the BS model parameters are considered. Panel (b) shows the 5 pairs of simultaneous (within 2 minutes) 
BS crossings of the Tianwen-1's orbiter and MAVEN. Each pair is indicated with the same colored dots, but the dots of Tianwen-1
are enclosed by black circles. The BS crossing time of the Tianwen-1's orbiter is given in 
the upper-right corner followed by a time interval with the positive value meaning the later BS crossing of MAVEN.
All these data are presented in the aberrated MSO coordinates.
}\label{fig:bscrossings}
\end{center}
\end{figure*}

\section{Bow Shock Crossings}\label{sec_bs}

Bow shock is one of the notable features in the Martian space environment. Its shape may reflect the upstream solar
wind conditions and solar EUV intensity and the interaction processes between solar wind and Martian 
atmosphere~\citep{Mazelle_etal_2004, Ramstad_etal_2017, Hall_etal_2019}. Thus, studying Martian BS is our first 
choice to show the science results of MOMAG. During 2021 November 13 -- December 31, we recognize 158 
BS crossings from MOMAG data by manually checking the magnetic field strength variation and the fluctuation level 
which is measured by the standard deviation of magnetic field within one minute. In principle, there 
should be more crossings, but Tianwen-1 mostly crossed the flank of the BS where the characteristic of a shock
may be too weak to be recognized. During the same period, we recognize 454 BS crossings from MAVEN/MAG data.
 
Figure~\ref{fig:bscrossings}a shows all of the BS crossings in the aberrated MSO coordinates (MSO coordinates are rotated 
by $4^\circ$ about the $z$-axis to reduce the effect of the Mars orbital motion on the solar wind flow direction).
Since the spatial coverage of the crossings is not wide enough, we do not try to fit these crossings to find the best-fit BS model, 
but instead to compare with the previously established BS model~\citep{Edberg_etal_2008}. We can find in the figure that the crossings
statistically match the model fairly well.

Martian BS position and global shape were derived from many single crossings. 
%and never checked by the simultaneously crossings of two probes. 
Now we can check this based on the joint magnetic field
observations from Tianwen-1/MOMAG and MAVEN/MAG. By assuming that the BS remains unchanged within 2 minutes,
we use two crossings of Tianwen-1 and MAVEN within 2 minutes to exam the BS global shape.  
We choose 2 minutes because the upstream solar wind conditions that determine the BS position and shape, i.e., 
the fast-mode Mach number and dynamic pressure, are usually stable within this time-scale as revealed by the 
following analysis.

\begin{figure*}[tbh]
\begin{center}
\includegraphics[width=0.7\hsize]{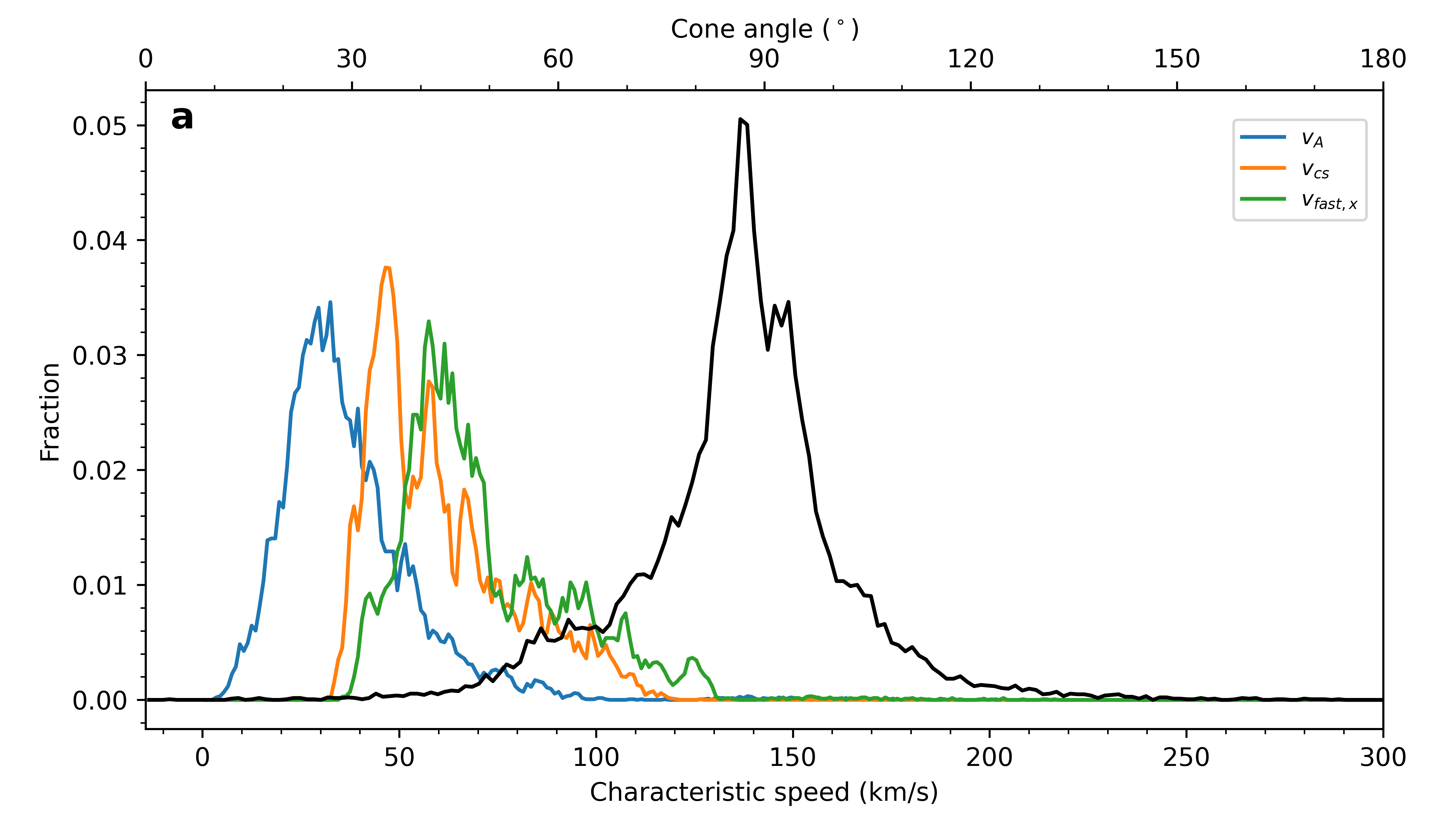}\\
\includegraphics[width=0.7\hsize]{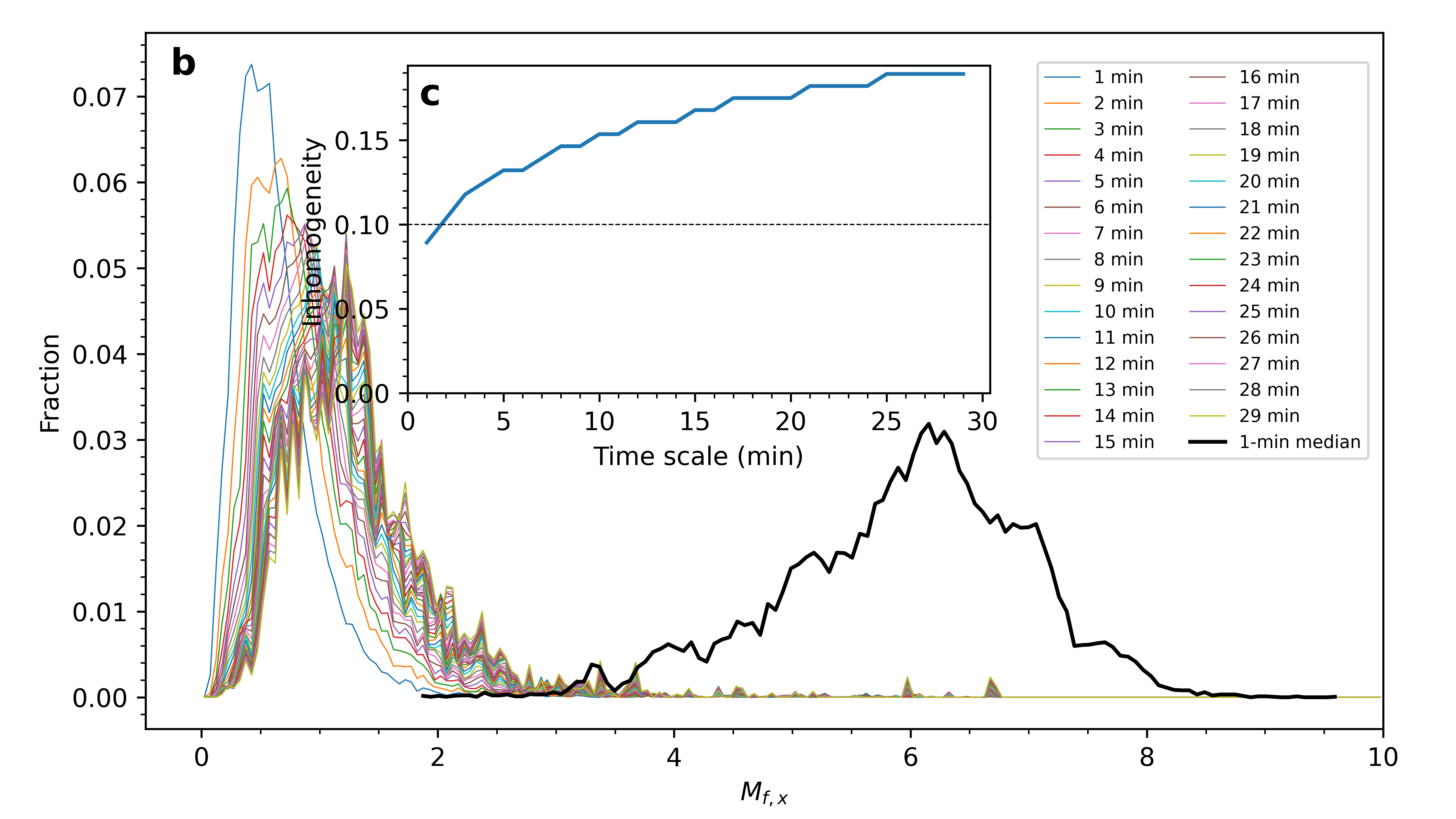}\\
\includegraphics[width=0.7\hsize]{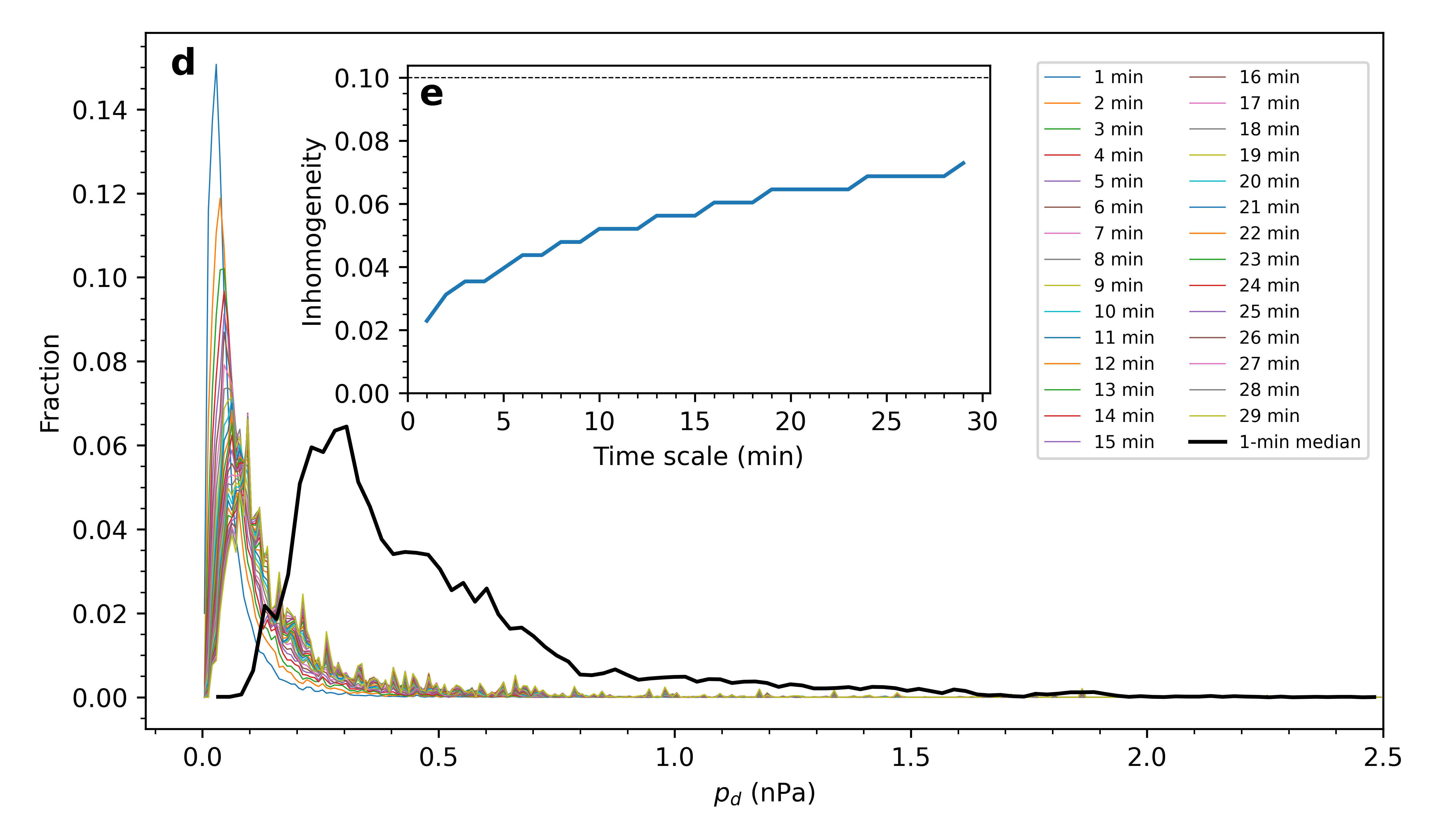}\\
	\caption{\footnotesize{Characteristic speeds and inhomogeneities of solar wind fast Mach number 
and dynamic pressure based on
MAVEN data. Panel (a) shows the distributions of the minutely-averaged Alfv{\'e}n speed (blue), sound speed (orange) and 
the fast-mode magnetoacoustic speed along the $x$ direction in MSO coordinates (green) 
during November 13 and December 31, 2021. 
The black line gives the distribution of the absolute value of the angle between the minutely-averaged magnetic field vector 
and the $x$ direction. Panel (b) shows the distributions of the range of the fast-mode Mach number
along the $x$ direction, $M_{f,x}$, within the various time scales, 
and the distribution of the minutely-averaged $M_{f,x}$ (see 
the main text for details). Panel (c) gives the inhomogeneity as a function of time scale. 
The definition of inhomogeneity here 
can be found in the main text. Panel (d) and (e) are for solar dynamic pressure 
with the same arrangement as Panel (b) and (c).}
}\label{fig:inhomogeneity}
\end{center}
\end{figure*}

Figure~\ref{fig:inhomogeneity}a shows the characteristic speeds in the solar wind, which are 
calculated every minute based on the MAVEN/SWIA~\citep{Halekas_etal_2017} measurements of the solar wind velocity,
ion density and temperature and MAVEN/MAG measurements of magnetic field during November 13 -- December 31, 2021.
The Alfv{\'e}n speed, $v_A$, ranges from almost zero to more than 100 km s$^{-1}$ with the peak around 30 km s$^{-1}$.
Since Alfv{\'e}n wave propagates along the magnetic field, if we take the direction of magnetic field, 
that mostly concentrates around 86$^\circ$ with respect to the $x$-axis in MSO (as indicated by the black 
line in Fig.\ref{fig:inhomogeneity}a), into account, the Alfv{\'e}n speed along the $x$-axis approaches zero. 
The sound speed, $v_{cs}$, is overall larger than the Alfv{\'e}n speed, and is rarely smaller than 30 km s$^{-1}$. 
The fast-mode magnetoacoustic speed along the $x$-axis, $v_{f,x}$, is overall larger than both the Alfv{\'e}n speed 
and the sound speed with its peak around 60 km s$^{-1}$. 

Since the solar wind propagates along the $x$-axis and $v_{f,x}$ is the fastest among these characteristic speeds, 
the fast-mode Mach number in $x$-axis, $M_{f,x}$, is calculated. \mod{The black line in Fig.\ref{fig:inhomogeneity}b shows
the median value of $M_{f,x}$ within one minute during the period of interest. We can read from the line that the
dynamic range of $M_{f,x}$ is about 7, i.e., ranging from about 2 to 9 with the peak at about 6.2.}
We further exam the inhomogeneity of $M_{f,x}$ by calculating the difference between 
the maximum and minimum values of $M_{f,x}$ within a given time scale varying from one minute to 29 minutes 
as shown by the color-coded thin lines in Figure~\ref{fig:inhomogeneity}b. Each line presents the distribution 
of the difference or the range of the $M_{f,x}$ in a given time scale. We can see that these distributions 
extend toward large values with increasing time scales, suggesting the enhancement of the inhomogeneity in terms of
$M_{f,x}$. Then we determine the middle value of $M_{f,x}$ for each distribution, at which the distribution is divided 
equally, and define the inhomogeneity as the ratio of the middle value to the dynamic range of $M_{f,x}$.
The dependence of the inhomogeneity on the time scale is plotted in Figure~\ref{fig:inhomogeneity}c. 
If considering that inhomogeneity of 0.1 is an acceptable level for a stale solar wind, we may conclude that
the time scale of stable solar wind is about 2 minutes in terms of $M_{f,x}$. 

The similar analysis is applied on the solar wind dynamic pressure, $p_d$, as shown in 
Figure~\ref{fig:inhomogeneity}d and e. The dynamic pressure also shows a single-peak distribution ranging 
from about 0.01 nPa to nearly 2.5 nPa with the peak around 0.3 nPa.
The inhomogeneity of $p_d$ also increases as the time scale increases. By setting the dynamic range of $p_d$ to
be 2, we find that the inhomogeneity is less than 0.1 even at the time scale of 30 minutes. 
This suggests that $M_{f,x}$ is much more dynamic than $p_d$ in the upstream of Martian BS.

Based on the above analysis, we search the BS-crossing pairs of Tianwen-1 and MAVEN within 
2 minutes in the period of interest, 
and plot the results in Figure~\ref{fig:bscrossings}b. A total of 5 pairs are found. 
A first impression is that the global shape
of the BS is slightly more flattened than the model. But this just reflects the south-north asymmetry of the Martian 
BS\citep[e.g.,][]{Edberg_etal_2008, Dubinin_etal_2008}, as Tianwen-1 orbiter crossed the southern flank of the BS, 
while MAVEN crossed the BS at low latitude on the northern hemisphere. 

\begin{figure*}[h]
\begin{center}
\includegraphics[width=\hsize]{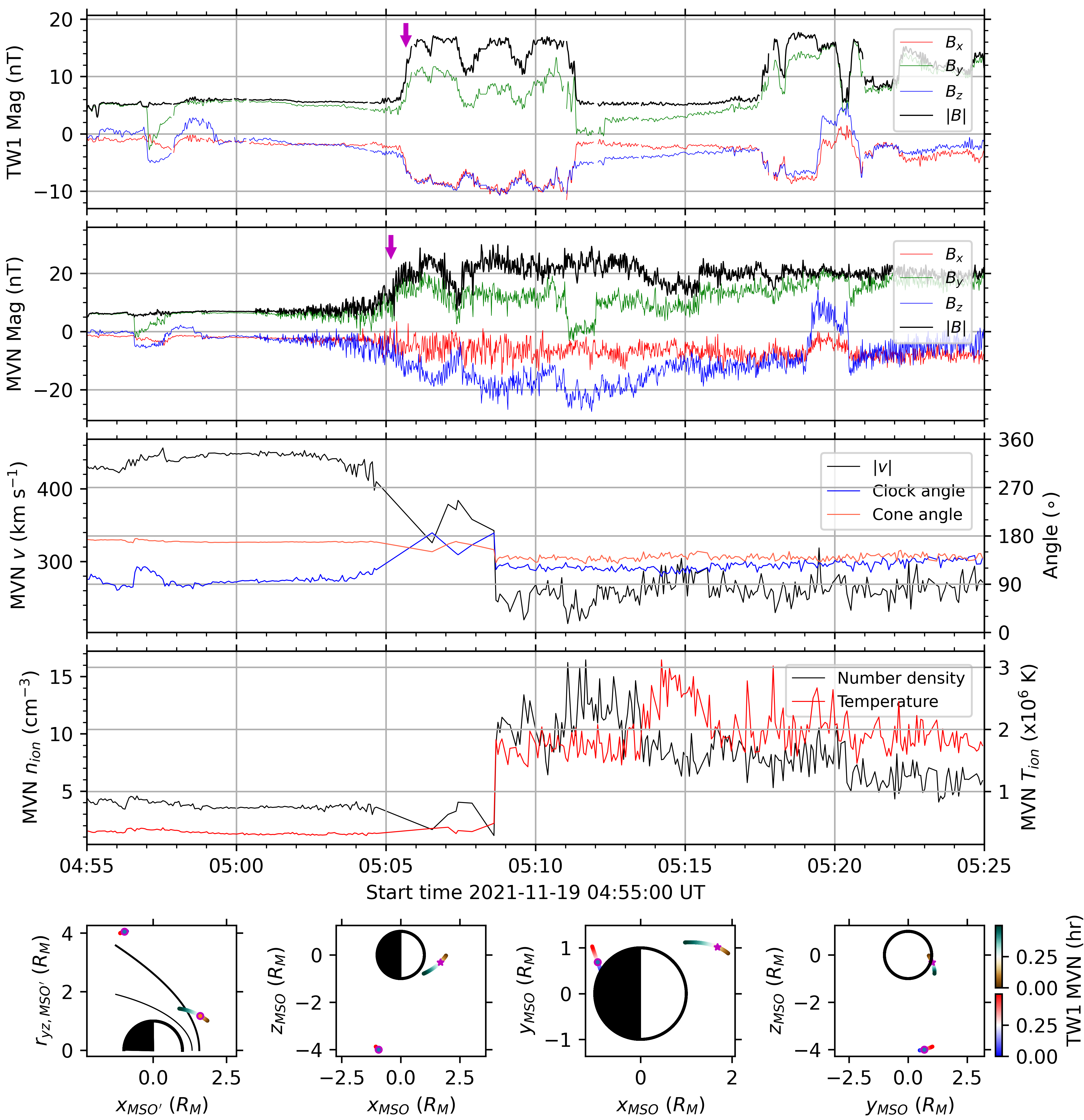}
	\caption{The simultaneous bow shock crossing around 05:05 UT on November 19, 2021. From the top to bottom, the panels
display the magnetic field measured by MOMAG, the magnetic field measured by MAVEN/MAG, and the solar wind velocity, 
the number density and temperature of ions measured by MAVEN/SWIA, and the orbits of Tianwen-1's orbiter and MAVEN viewed from 
different angles. The purple arrows in the first two panels indicate the times of the bow shock crossings, and 
the markers in the panels on the bottom indicate the positions of the crossings.
}\label{fig:bs1}
\end{center}
\end{figure*}

\begin{figure*}[h]
\begin{center}
\includegraphics[width=\hsize]{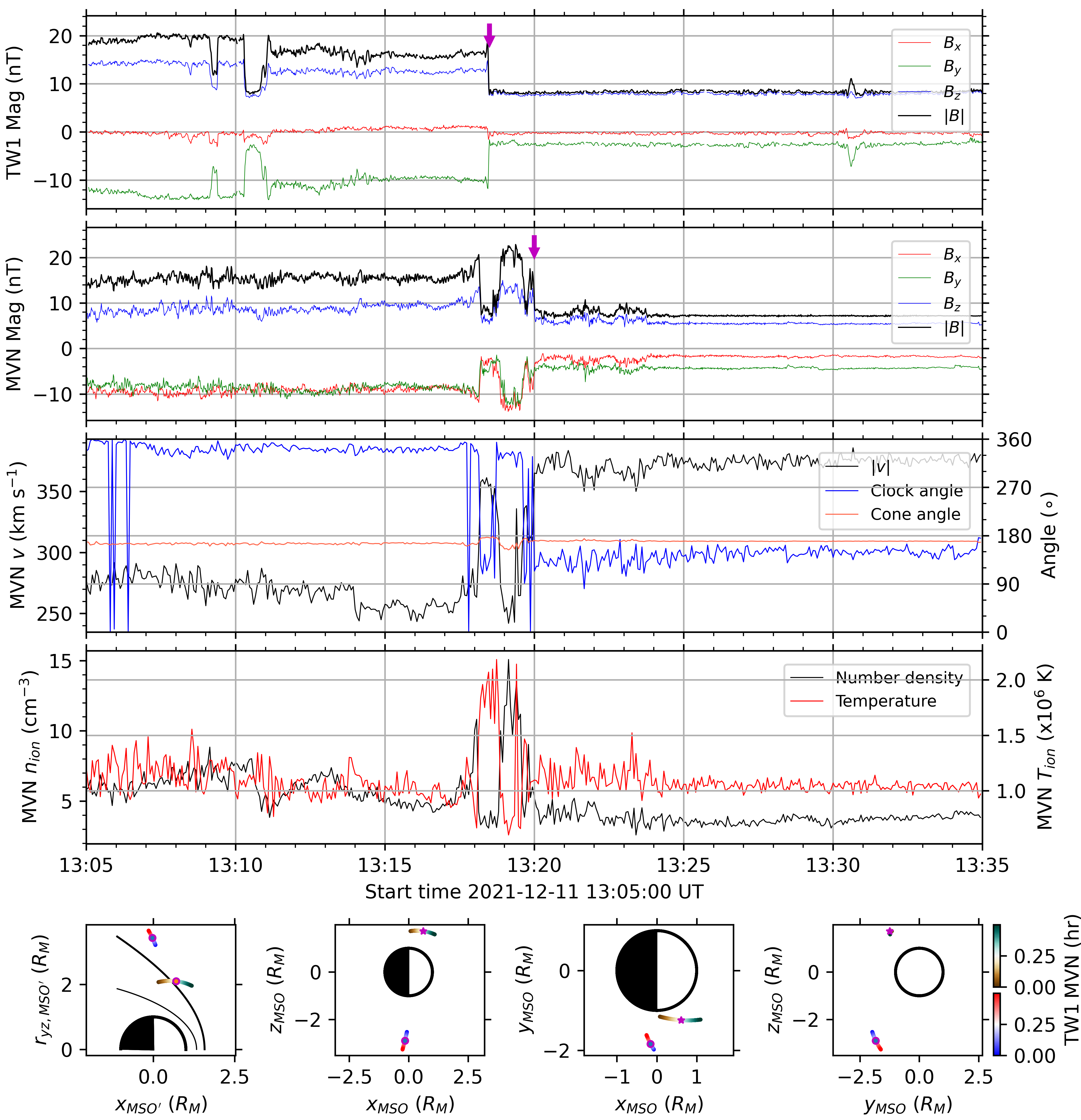}
	\caption{The simultaneous bow shock crossing around 13:19 UT on December 11, 2021.
}\label{fig:bs2}
\end{center}
\end{figure*}

\begin{figure*}[h]
\begin{center}
\includegraphics[width=\hsize]{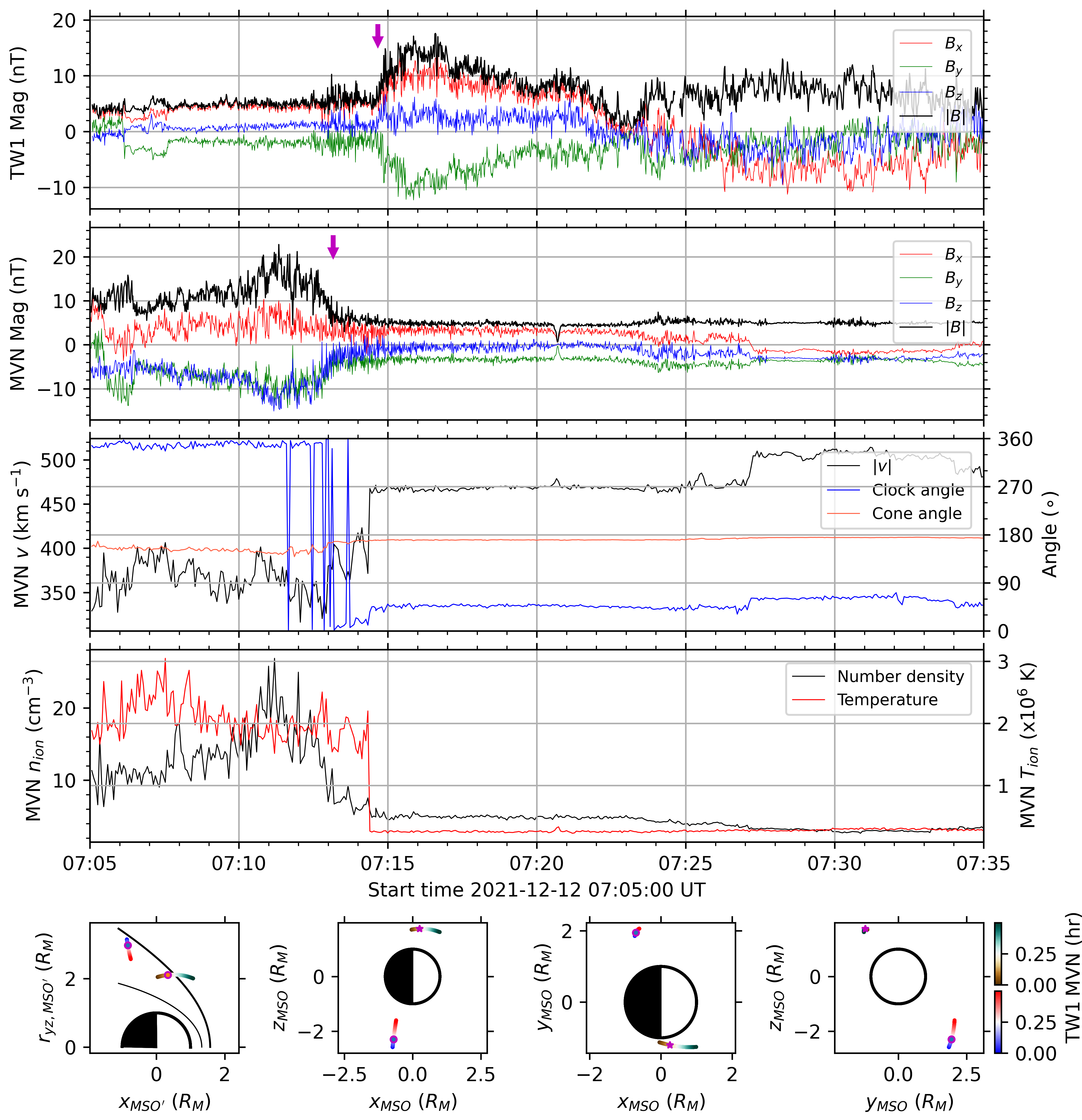}
	\caption{The simultaneous bow shock crossing around 07:14 UT on December 12, 2021.
}\label{fig:bs3}
\end{center}
\end{figure*}

\begin{figure*}[h]
\begin{center}
\includegraphics[width=\hsize]{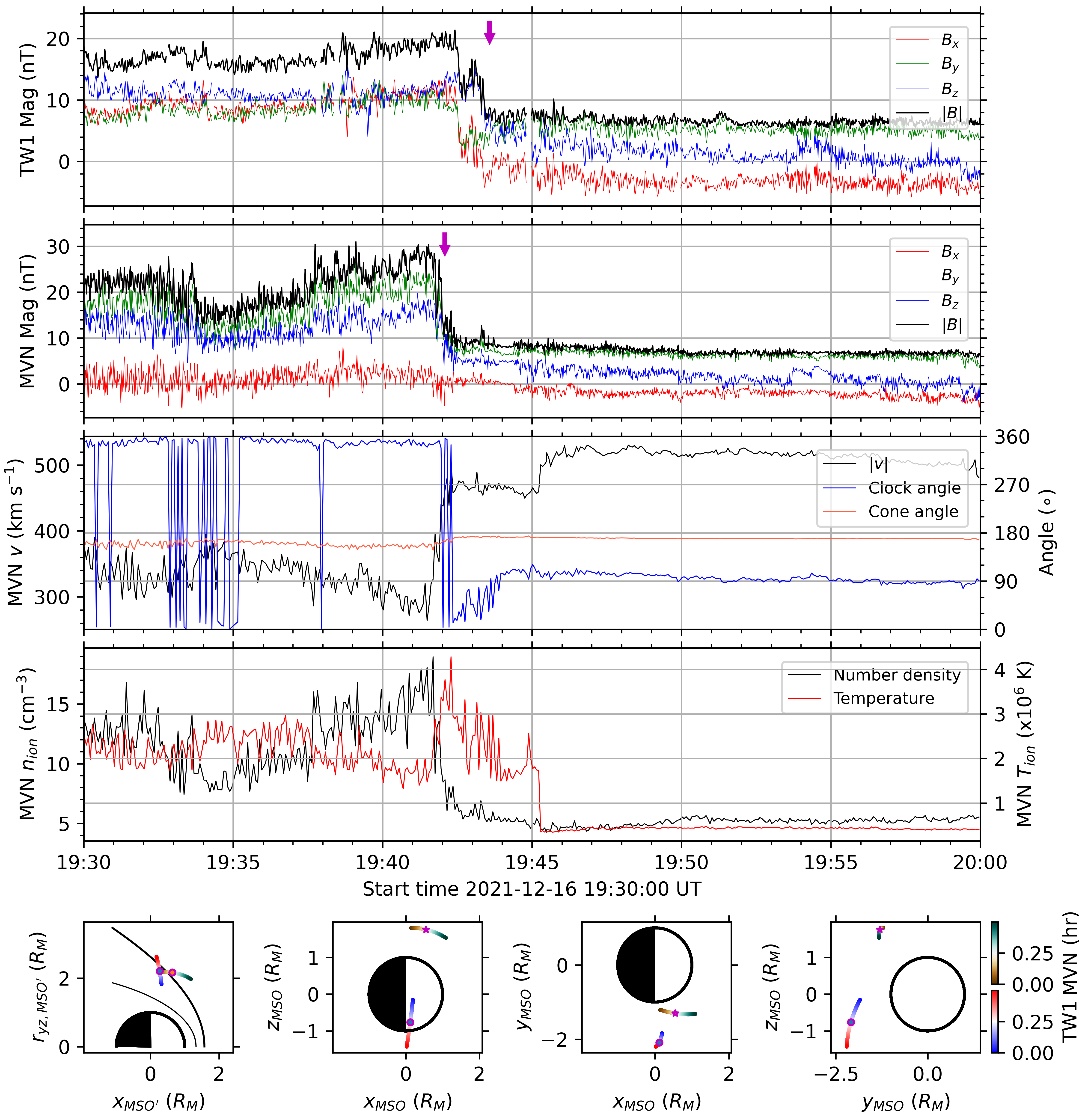}
	\caption{The simultaneous bow shock crossing around 19:43 UT on December 16, 2021.
}\label{fig:bs4}
\end{center}
\end{figure*}

\begin{figure*}[h]
\begin{center}
\includegraphics[width=\hsize]{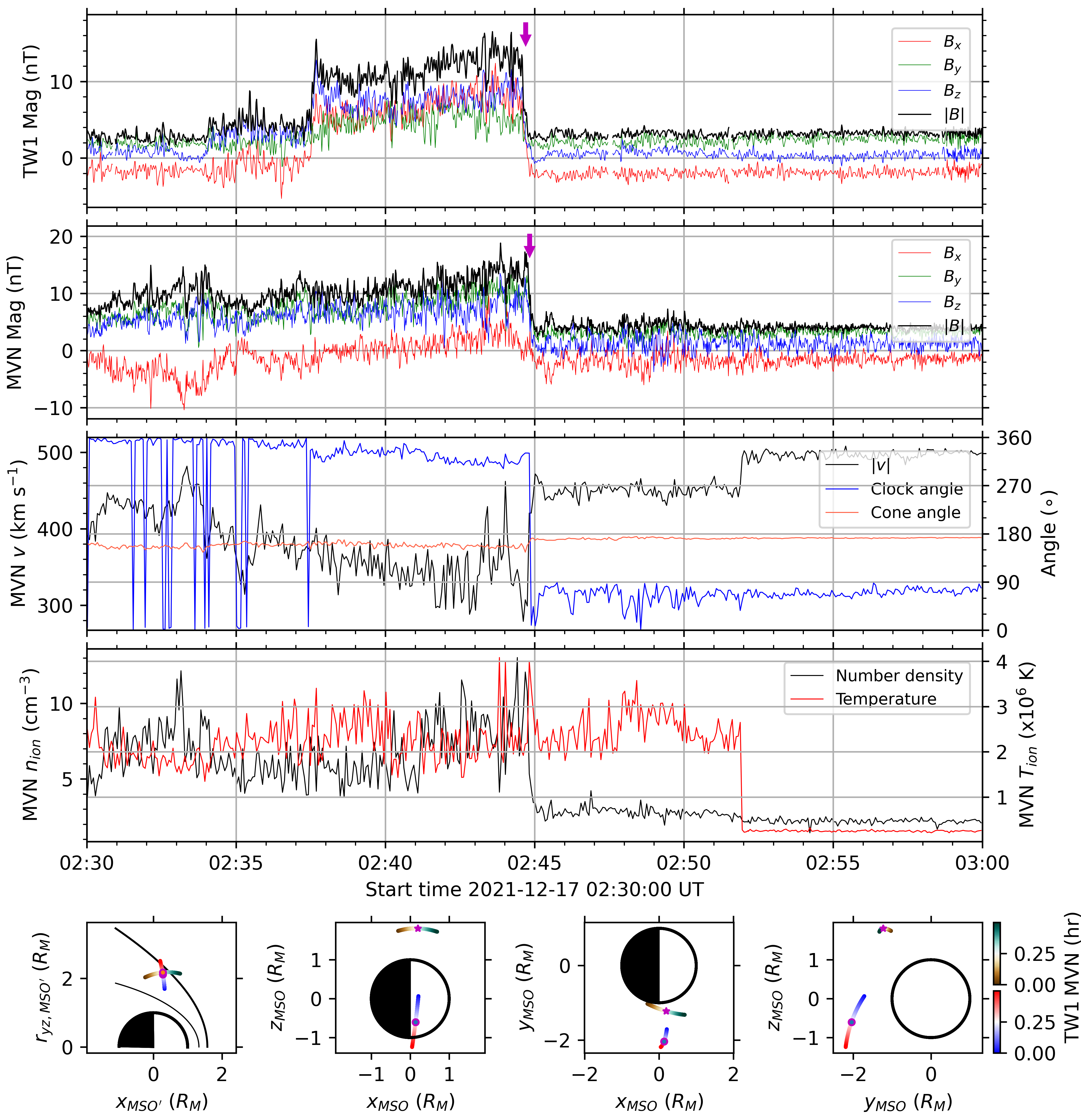}
	\caption{The simultaneous bow shock crossing around 02:44 UT on December 17, 2021.
}\label{fig:bs5}
\end{center}
\end{figure*}

Figures~\ref{fig:bs1}--\ref{fig:bs5} show the 5 pairs of the BS crossings. Around 05:05 UT on November 19, both spacecraft
crossed the BS from the solar wind into magnetosheath, when Tianwen-1 orbiter was far above the southern pole of Mars and 
MAVEN was close to the BS nose (see Fig.\ref{fig:bs1}). The magnetic fields in the solar wind measured before they 
entered magnetosheath look 
quite similar. Between 04:56 and 05:00 UT, we can see the large variation patterns in the three components of the 
magnetic fields without a significant change in the total magnitude, which are probably the features of an Alfv\'en wave. 
This featured structure arrived at Tianwen-1 orbiter later than MAVEN by nearly 30 s, which was roughly the time spent by 
solar wind travelling from MAVEN to Tianwen-1. 

The magnetic fields measured by the two spacecraft after they crossed the BS 
show different patterns.  
From the first panel of Figure~\ref{fig:bs1}, it seems that the Tianwen-1 orbiter crossed the BS three times within 7 minutes,
finally returned back to solar wind at 05:11:30 UT, and at about 05:17:20 UT, the orbiter started to cross the BS again.
Not like Tianwen-1, MAVEN stayed in the magnetosheath after the crossing except one turning back at around 05:07:30 UT. 
\mod{It is similar to the case shown in Figure~\ref{fig:obstw1}h and \ref{fig:obsmvn}h, of which the Tianwen-1 
orbiter crossed the BS three times in 7 minutes, but MAVEN had only one clear crossing.}
These phenomena suggest that the Martian BS is very dynamic with the time scale even less than one minute, and the BS flank 
is more dynamic than the nose during this time period. 
Such multiple-crossings in minutes deserve further study, especially for events with Tianwen-1
orbiter crossing the BS and MAVEN staying in the solar wind to monitor the upstream condition.

The second pair of the BS crossings is found around 13:19 UT on December 11 as shown in Figure~\ref{fig:bs2}. 
Both spacecraft were crossing the BS from magnetosheath to the solar wind. We can see a sharp jump at 13:18:30 UT in the
MOMAG data, and a sharp jump at 13:20:00 UT in the MAVEN/MAG data. In both data, we also can find a large dip
in the total magnetic field strength. It is hard to determine if they are correlated. The third pair was around
07:14 UT on December 12 with one crossing from the solar wind into magnetosheath and the other from magnetosheath into
the solar wind (Fig.\ref{fig:bs3}). The fourth pair was around 19:43 UT on December 16. Both spacecraft travelled
from the magnetosheath into the solar wind (Fig.\ref{fig:bs4}). The last pair is found around 02:45 UT on December
17. Both spacecraft also travelled from the magnetosheath into the solar wind (Fig.\ref{fig:bs5}). If looking at 
the total strengths of the magnetic fields in the magnetosheath for all the BS crossing pairs, we may find they 
are more or less similar no matter how large in distance the two spacecraft are apart, suggesting the 
consistency of the global magnetic structure surrounding Mars at the level of large scale.

\newpage

\newpage

\section{Summary}\label{sec_sum}

we have presented the in-flight performance and first results of Tianwen-1/MOMAG with the focus on the most notable structure 
--- Martian BS. Based on the first one and a half months' data, we identified 158 clear BS crossings, whose 
locations are consistent with the BS model in statistics. The simultaneous 
BS crossings of the Tianwen-1 and MAVEN verified the south-north asymmetry of the BS, and also showed the 
similarity of magnetic field profiles from the two spacecraft. The first pair of the simultaneous BS crossings \mod{along
with the BS crossing case on December 30} suggests that
the BS is probably more dynamic at flank than near the nose.
By comparing with the MAVEN observations, we also found similar structures propagating with 
the solar wind from MAVEN to the Tianwen-1 orbiter. 
We conclude that MOMAG shows an excellent performance and provides accurate measurements of magnetic field vectors. 
Now MOMAG has scanned the magnetic field in the MPR, magnetosheath and solar wind near the dawn-dusk side. These measurements
along with the MAVEN data will help us better understand the plasma environment surrounding Mars.

\begin{acknowledgments}
We acknowledge the use of the data from the MAG and SWIA onboard MAVEN spacecraft, which are obtained from 
NASA Planetary Data System (\url{https://pds-ppi.igpp.ucla.edu/}). One may apply for the Tianwen-1/MOMAG data at CNSA
Data Release System (\url{http://202.106.152.98:8081/marsdata/}) or can just download the data used in this paper 
from the official website of the MOMAG team (\url{http://space.ustc.edu.cn/dreams/tw1_momag/}).
The work is support by the NSFC (Grant Nos 42130204 and 42188101) and
the Strategic Priority Program of the Chinese Academy of Sciences (Grant No. XDB41000000).
%We also thank for the support from National Space Science Data Center. 
Y.W. is particularly grateful to the support of the Tencent Foundation.
\end{acknowledgments}

\bibliographystyle{agu}
\bibliography{../../ahareference}

\begin{thebibliography}{23}
\providecommand{\natexlab}[1]{#1}
\expandafter\ifx\csname urlstyle\endcsname\relax
  \providecommand{\doi}[1]{doi:\discretionary{}{}{}#1}\else
  \providecommand{\doi}{doi:\discretionary{}{}{}\begingroup
  \urlstyle{rm}\Url}\fi

\bibitem[{\textit{Bertucci et~al.}(2004)}]{Bertucci_etal_2004}
Bertucci, C., et~al., {MGS MAG/ER} observations at the magnetic pileup boundary
  of mars: Draping enhancement and low frequency waves, \textit{Adv. Space
  Res.}, \textit{33}, 1938--1944, \doi{10.1016/j.asr.2003.04.054}, 2004.

\bibitem[{\textit{Brain et~al.}(2006)}]{Brain_etal_2006}
Brain, D.~A., et~al., On the origin of aurorae on {Mars}, \textit{Geophys. Res.
  Lett.}, \textit{33}, \doi{10.1029/2005GL024782}, 2006.

\bibitem[{\textit{Brain et~al.}(2015)}]{Brain_etal_2015}
Brain, D.~A., et~al., The spatial distribution of planetary ion fluxes near
  {Mars} observed by {MAVEN}, \textit{Geophys. Res. Lett.}, \textit{42},
  9142--9148, \doi{10.1002/2015GL065293}, 2015.

\bibitem[{\textit{Connerney et~al.}(2015)\textit{Connerney, Espley, Lawton,
  Murphy, Odom, Oliversen, and Sheppard}}]{Connerney_etal_2015a}
Connerney, J.~E., J.~Espley, P.~Lawton, S.~Murphy, J.~Odom, R.~Oliversen, and
  D.~Sheppard, {The {MAVEN} magnetic field investigation}, \textit{Space
  Science Reviews}, \textit{195}, 257--291, \doi{10.1007/s11214-015-0169-4},
  2015.

\bibitem[{\textit{Dubinin et~al.}(2008)}]{Dubinin_etal_2008}
Dubinin, E., et~al., Structure and dynamics of the solar wind/ionosphere
  interface on {Mars}: {MEX-ASPERA-3} and {MEX-MARSIS} observations,
  \textit{Geophys. Res. Lett.}, \textit{35}, \doi{10.1029/2008GL033730}, 2008.

\bibitem[{\textit{Edberg et~al.}(2008)\textit{Edberg, Lester, Cowley, and
  Eriksson}}]{Edberg_etal_2008}
Edberg, N. J.~T., M.~Lester, S.~W.~H. Cowley, and A.~I. Eriksson, Statistical
  analysis of the location of the {Martian} magnetic pileup boundary and bow
  shock and the influence of crustal magnetic fields, \textit{J. Geophys. Res.:
  Space Phys.}, \textit{113}, A08,206, \doi{10.1029/2008JA013096}, 2008.

\bibitem[{\textit{Ergun et~al.}(2006)\textit{Ergun, Andersson, Peterson, Brain,
  Delory, Mitchell, Lin, and Yau}}]{Ergun_etal_2006}
Ergun, R.~E., L.~Andersson, W.~K. Peterson, D.~Brain, G.~T. Delory, D.~L.
  Mitchell, R.~P. Lin, and A.~W. Yau, Role of plasma waves in {Mars}'
  atmospheric loss, \textit{Geophys. Res. Lett.}, \textit{33},
  \doi{10.1029/2006GL025785}, 2006.

\bibitem[{\textit{Halekas et~al.}(2017)}]{Halekas_etal_2017}
Halekas, J.~S., et~al., Structure, dynamics, and seasonal variability of the
  {Mars}-solar wind interaction: {MAVEN} solar wind ion analyzer in-flight
  performance and science results, \textit{J. Geophys. Res.: Space Phys.},
  \textit{122}, 547--578, \doi{10.1002/2016JA023167}, 2017.

\bibitem[{\textit{Hall et~al.}(2019)\textit{Hall, S{\'{a}}nchez-Cano, Wild,
  Lester, and Holmstr{\"{o}}m}}]{Hall_etal_2019}
Hall, B.~E., B.~S{\'{a}}nchez-Cano, J.~A. Wild, M.~Lester, and
  M.~Holmstr{\"{o}}m, {The Martian Bow Shock Over Solar Cycle 23–24 as
  Observed by the {Mars Express} Mission}, \textit{J. Geophys. Res.: Space
  Phys.}, \textit{124}, 4761--4772, \doi{10.1029/2018JA026404}, 2019.

\bibitem[{\textit{Jakosky et~al.}(2015{\natexlab{a}})}]{Jakosky_etal_2015}
Jakosky, B.~M., et~al., The mars atmosphere and volatile evolution ({MAVEN})
  mission, \textit{Space Sci. Rev.}, \textit{195}, 3--48, 2015{\natexlab{a}}.

\bibitem[{\textit{Jakosky et~al.}(2015{\natexlab{b}})}]{Jakosky_etal_2015a}
Jakosky, B.~M., et~al., {MAVEN} observations of the response of {Mars} to an
  interplanetary coronal mass ejection, \textit{Science}, \textit{350},
  aad0210, \doi{10.1126/science.aad0210}, 2015{\natexlab{b}}.

\bibitem[{\textit{Liu et~al.}(2020)}]{LiuK_etal_2020}
Liu, K., et~al., Mars orbiter magnetometer of {China}'s first mars mission
  {Tianwen}-1, \textit{Earth \& Planet. Phys.}, \textit{4}, 384--389, 2020.

\bibitem[{\textit{Mazelle et~al.}(2004)}]{Mazelle_etal_2004}
Mazelle, C., et~al., Bow shock and upstream phenomena at {Mars}, \textit{Space
  Sci. Rev.}, \textit{111}, 115--181, \doi{10.1023/B:SPAC.0000032717.98679.d0},
  2004.

\bibitem[{\textit{Pope et~al.}(2011)\textit{Pope, Zhang, Balikhin, Delva,
  Hvizdos, Kudela, and Dimmock}}]{Pope_etal_2011}
Pope, S., T.~Zhang, M.~Balikhin, M.~Delva, L.~Hvizdos, K.~Kudela, and
  A.~Dimmock, Exploring planetary magnetic environments using magnetically
  unclean spacecraft: A systems approach to {VEX MAG} data analysis,
  \textit{Ann. Geophys.}, \textit{29}, 639--647,
  \doi{10.5194/angeo-29-639-2011}, 2011.

\bibitem[{\textit{Ramstad et~al.}(2017)\textit{Ramstad, Barabash, Futaana, and
  Holmstr{\"{o}}m}}]{Ramstad_etal_2017}
Ramstad, R., S.~Barabash, Y.~Futaana, and M.~Holmstr{\"{o}}m, Solar wind- and
  {EUV}-dependent models for the shapes of the {Martian} plasma boundaries
  based on {Mars Express} measurements, \textit{J. Geophys. Res.: Space Phys.},
  \textit{122}, 7279--7290, \doi{10.1002/2017JA024098}, 2017.

\bibitem[{\textit{Ruhunusiri et~al.}(2017)}]{Ruhunusiri_etal_2017}
Ruhunusiri, S., et~al., Characterization of turbulence in the mars plasma
  environment with {MAVEN} observations, \textit{J. Geophys. Res.: Space
  Phys.}, \textit{122}, 656--674, \doi{10.1002/2016JA023456}, 2017.

\bibitem[{\textit{Russell and Blanco-Cano}(2007)}]{Russell_Blanco-Cano_2007}
Russell, C.~T., and X.~Blanco-Cano, Ion-cyclotron wave generation by planetary
  ion pickup, \textit{J. Atmos. Solar-Terres. Phys.}, \textit{69}, 1723--1738,
  2007.

\bibitem[{\textit{Wan et~al.}(2020)\textit{Wan, Wang, Li, and
  Wei}}]{WanW_etal_2020}
Wan, W.~X., C.~Wang, C.~L. Li, and Y.~Wei, China's first mission to {Mars},
  \textit{Nat. Astron.}, \textit{4}, 721--721, \doi{10.1038/s41550-020-1148-6},
  2020.

\bibitem[{\textit{Wang and Pan}(2021)}]{WangG_etal_2021}
Wang, G., and Z.~Pan, A new method to calculate the fluxgate magnetometer
  offset in the interplanetary magnetic field, \textit{J. Geophys. Res.: Space
  Phys.}, \textit{126}, \doi{10.1029/2020JA028893}, 2021.

\bibitem[{\textit{Zhang et~al.}(2008)}]{ZhangT_etal_2008}
Zhang, T.~L., et~al., Initial {Venus Express} magnetic field observations of
  the {Venus} bow shock location at solar minimum, \textit{Planet. Space Sci.},
  \textit{56}, 785--789, \doi{10.1016/j.pss.2007.09.012}, 2008.

\bibitem[{\textit{Zhang et~al.}(2012)}]{ZhangT_etal_2012}
Zhang, T.~L., et~al., Magnetic reconnection in the near {Venusian} magnetotail,
  \textit{Science}, \textit{336}, 567--570, \doi{10.1126/science.1217013},
  2012.

\bibitem[{\textit{Zou et~al.}(2021)}]{ZouY_etal_2021}
Zou, Y., et~al., Scientific objectives and payloads of {Tianwen-1}, {China}'s
  first {Mars} exploration mission, \textit{Adv. Space Res.}, \textit{67},
  812--823, 2021.

\bibitem[{\textit{Zou et~al.}(2023)}]{ZouZ_etal_2022}
Zou, Z., et~al., In-flight calibration of the magnetometer on the mars orbiter
  of tianwen-1, \textit{Sci. China Tech. Sci.}, \textit{submitted}, 2023.

\end{thebibliography}

\end{article}
\end{document}